\documentclass[a4paper]{article}

\usepackage{INTERSPEECH2020}
\usepackage{amsmath,graphicx}
\usepackage{import}
\usepackage[utf8]{inputenc}
\usepackage{url}
\usepackage{hhline}
\usepackage{tabularx}
\usepackage{makecell}
\usepackage{multicol}
\usepackage{multirow}
\usepackage{subcaption}
\usepackage{graphicx}
\usepackage{hyperref}
\usepackage{pgfplots}
\usepackage{array}
\usepackage{xcolor}
\usepackage{commath}
\usepackage{booktabs,ragged2e}

\newcommand\blfootnote[1]{%
  \begingroup
  \renewcommand\thefootnote{}\footnote{#1}%
  \addtocounter{footnote}{-1}%
  \endgroup
}

\newcolumntype{L}[1]{>{\RaggedRight\arraybackslash}p{#1}}
\newcolumntype{R}[1]{>{\RaggedLeft\arraybackslash}p{#1}}
\newcolumntype{C}[1]{>{\centering\arraybackslash}p{#1}}
\newcommand\doubleplus{+\kern-1.3ex+\kern0.8ex}
\renewcommand{\vec}[1]{\mathbf{#1}}

\title{FaceFilter: Audio-visual speech separation using still images}
\name{Soo-Whan Chung$^{1}$, Soyeon Choe$^{2}$,  Joon Son Chung$^{2}$ and Hong-Goo Kang$^{1}$}
\address{
$^{1}$Department of Electrical \& Electronic Engineering, Yonsei University, Seoul, South Korea\\
$^{2}$Naver Corp., Seongnam-si, Gyeonggi-do, South Korea}
\email{\vspace{-13pt}} 

\begin{document}

\maketitle
\begin{abstract}
The objective of this paper is to separate a target speaker's speech from a mixture of two speakers using a deep audio-visual speech separation network.
Unlike previous works that used lip movement on video clips or pre-enrolled speaker information as an auxiliary conditional feature, we use a single face image of the target speaker.
In this task, the conditional feature is obtained from facial appearance in cross-modal biometric task, where audio and visual identity representations are shared in latent space.
Learnt identities from facial images enforce the network to isolate matched speakers and extract the voices from mixed speech.
It solves the permutation problem caused by swapped channel outputs, frequently occurred in speech separation tasks.
The proposed method is far more practical than video-based speech separation since user profile images are readily available on many platforms. 
Also, unlike speaker-aware separation methods, it is applicable on separation with unseen speakers who have never been enrolled before.
We show strong qualitative and quantitative results on challenging real-world examples. 
 
\end{abstract}


\section{Introduction}
\label{sec:intro}
\blfootnote{\hspace{-12pt} Video examples: \url{https://youtu.be/ku9xoLh62E4}}

In recent years, there has been great progress in the field of automatic speech recognition, achieving human-level transcription accuracy in quiet environment~\cite{chan2016las,amodei2016deepspeech2,chiu2018s2s}.
However, it still remains a challenge to accurately recognise speech in noisy environments or multi-talker background.
Humans have a remarkable ability in separating one person's voice from others~\cite{mesgarani2012selective}.
In a {\em cocktail party} environment, for example, humans can focus on one speaker's voice while filtering out the rest. 

The goal of this work is to extract a target speaker's speech from a mixed signal. While there have been researches on separating simultaneous speakers based only on the audio~\cite{hershey2016deep,luo2018tasnet}, the permutation problem remains unsolved.
Although a permutation invariant training method~\cite{yu2017pit,kolbaek2017upit} relieves the problem, the permutation problem still exists in the inference stage since there is no explicit constraint for the channel assignment.
Furthermore~\cite{Wang2019voicefilter,xu2019tsenet} shows that an embedding of the target speaker’s voice can be used to separate simultaneously speaking speech as auxiliary information.
However, this requires a pre-enrollment of speaker embedding in a clean environment, which also limits its applications. 

In recent works of~\cite{Afouras18,Ephrat18,lu2019avdc,ochiai2019multimodal,Owens18}, it has been shown that the use of video helps in solving the cocktail party problem.
The methods are able to separate speech of a particular speaker by conditioning on his/her lip movements in the corresponding face video.
These audio-visual models have demonstrated impressive results, but the dependence on the lip movements means that their usefulness is limited to scenarios where high frame rate video is available.

But what if a single face image can substitute the full video stream?
This can be used in a wide range of scenarios since users' profile images are already available in many mobile devices, social networks and company groupware.
Once considered a seemingly impossible task, recent works have shown that voice representations can be obtained from face images and vice-versa~\cite{Nagrani18,Nagrani18c,chung2020pm,oh2019speech2face,kim2018learning}.

We propose a novel audio-visual speech separation (AVSS) model, \textit{FaceFilter}, that is conditioned on a still face image of a target speaker.
With our approach, the face image provides speaker identity information to the network, thus it resolves the ambiguity of assigning the separated voice to the speaker.
We make the following contributions: 
(i) we propose a novel method for audio-visual speech separation using only a single face image of the target speaker. The method exploits cross-modal biometric representation obtained from face appearance to solve the permutation problem in speech separation; 
(ii) the model trained on a large-scale dataset is evaluated on unseen and unheard speakers, on which we demonstrate strong qualitative and quantitative performance;
(iii) we conduct further experiments with additional methods for improving separation performance such as temporal attention~\cite{lin2017selfatt} and speaker representation loss~\cite{mun2020sound}.

\begin{figure*}[t]
    \centering
    \fbox{
    \begin{minipage}{2\columnwidth}
        \includegraphics[width=\columnwidth]{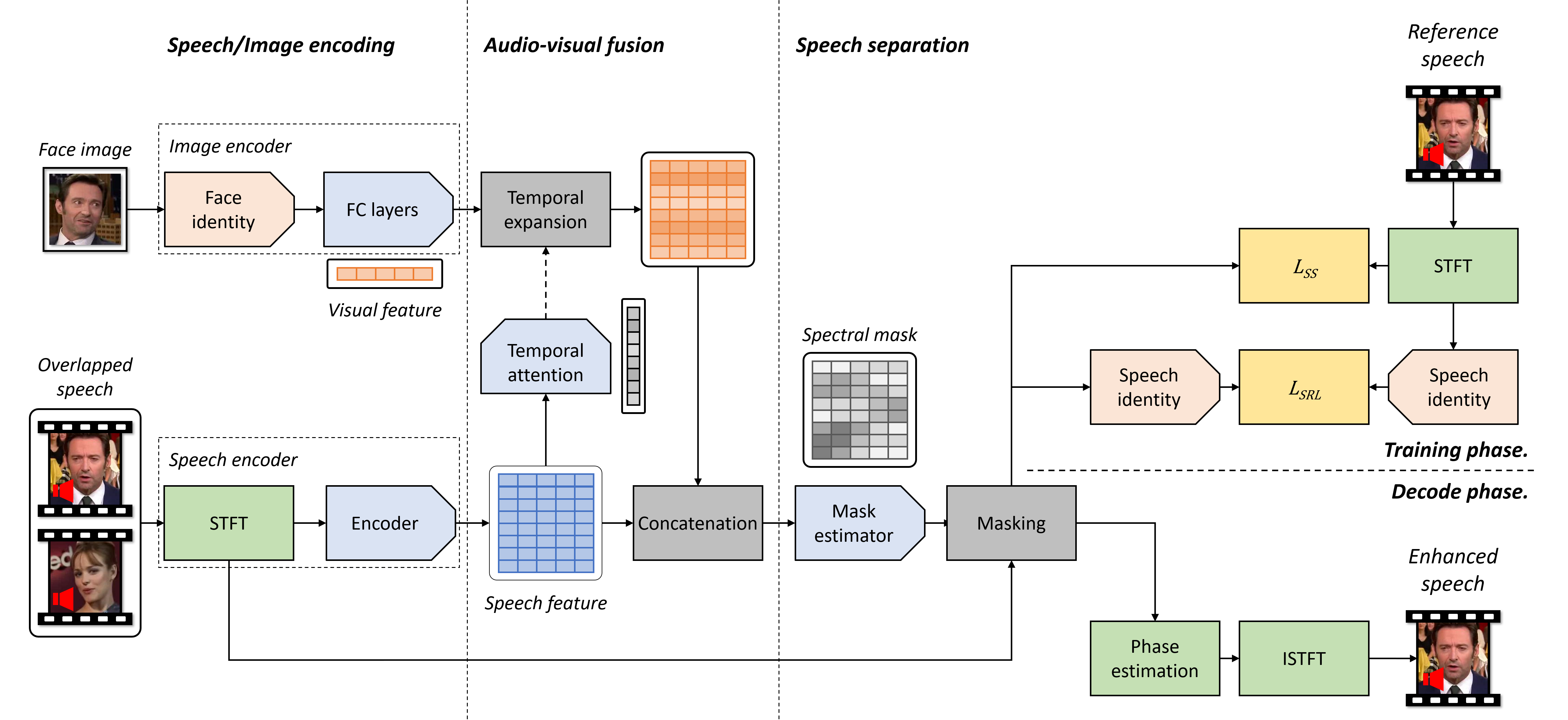}
    \end{minipage}
    }
    \caption{Description of the proposed audio-visual speech separation network. Blue blocks are trainable neural networks whereas the red blocks are pre-trained in cross-modal biometric task.}
    \label{fig:avsearc}
    \vspace{-6pt}
\end{figure*}

\section{Related works}
Research in speech separation has seen significant progress due to the application of neural network models and the availability of new large-scale datasets~\cite{Chung18a,panayotov2015librispeech}.
The majority of works on speech separation use the {\em audio} modality only, and they focus on how to effectively extract the portion of a target speaker by analysing spectral characteristic.
Spectral embeddings~\cite{hershey2016deep,luo2018danet} and time-domain speech analysis~\cite{luo2018tasnet,luo2019convtasnet} have shown great success on multiple speaker separation tasks by distinguishing one speaker's voice from mixed input speech.
However, they emit separated signals into random channels on the inference stage, where it requires additional clustering step that is vulnerable to permutation.
Some works have experimented on the use of speaker embeddings as conditional vectors to regulate the speech assignment.
In~\cite{Wang2019voicefilter,xu2019tsenet,vzmolikova2019speakerbeam}, they obtained speaker embeddings of target speaker from a pre-trained speaker representation model such as x-vector or i-vector~\cite{snyder2018xvector,dehak2010ivector}, and then used the representation as a conditional vector to extract the target speaker's voice from a noisy input speech signal.
The performance of these methods far exceeds the previous methods not conditioned on speaker-specific information.

There have been some works that use auxiliary audio-visual information, namely lip movements on video~\cite{Afouras18, Ephrat18, Gabbay2018, afouras2019my}.
In these methods, visual embedding learns linguistic information in a synchronisation task which matches natural correspondence between speech and visual streams.
They effectively extract target voice, even with a large number of simultaneous speech;
however, these methods are very sensitive to the quality of video.

\section{Cross-modal identity representation}
\label{sec:background}

Face and voice analysis are popular tools to verify user identity due to their non-invasive nature and good performance.
Although the two types of signals share common information ({\em i.e.} person identity), researches in each field have evolved independently~\cite{snyder2018xvector,schroff2015facenet} due to the difficulty in building a unified statistical model using heterogeneous features from the different modalities.
Speaker recognition analyses consistent spectral characteristics over utterances such as vocal tract shape, pitch and prosody variation, whereas face recognition looks into the shape of facial landmarks.
However, there is common higher-order information that are captured by both tasks, including but not limited to age, nationality and gender.

Recent works have demonstrated that joint embedding of face and voice characteristics can be learnt using deep neural networks trained on large-scale datasets.
In~\cite{Nagrani18,chung2020pm,Nagrani20d}, the problem is set up as a cross-modal matching task using a two stream network.
Representations of speech segments and facial appearances are mapped using the networks onto a joint embedding space, and they are trained to predict whether or not the pair comes from same identity.
The embeddings are learnt in a self-supervised manner; 
if audio segment and face image are taken from same video clip, the learning criterion minimises the distance between embeddings whereas it is maximised if they are from different clips.

We exploit these cross-modal identity embeddings to retrieve speaker identity from facial appearance.
In particular, we use the implementation of \cite{chung2020pm} in this paper, where the embeddings are trained as a one-of-many matching task.
We sample images and audio segments from talking face videos and train the model with a multi-way matching objective~\cite{chung2020pm}, where the model has to select one of the 200 audio segments that comes from the same video as the face image.
Both audio and video streams are based on the VGG-M network~\cite{Chatfield14} -- detailed network architecture is given in~\cite{chung2020pm}.

\section{Audio-visual speech separation}
\label{sec:proposed}
In this section, we describe the proposed architecture and training strategies for the AVSS network.

The most relevant work to the proposed method is the speaker-conditioned speech separation tasks.
In this paper, we replace the speaker identity vector with cross-modal identity embeddings extracted from face images instead of that from pre-enrolled speakers' voices.
Without having a speaker pre-enrollment step, we can retrieve speaker identity from a profile image in the inference stage, even for {\em unseen} speaker's voice.
The network structure that we use is similar to the one proposed in~\cite{Afouras18}, but with a few changes.
It consists of three sub-blocks such as speech/image encoders, audio-visual fusion, and speech separation.
The overall structure is illustrated in Figure~\ref{fig:avsearc}.

\vspace{2pt}
\noindent{\bf Speech/Image encoder.}
In this stage, speech and face appearances are extracted and encoded onto a latent space, which prepares for the audio-visual fusion.
The image encoder block extracts face representations to condition mask estimator.
The face identity extractor is pre-trained with the cross-modal identity matching task (Section~\ref{sec:background}), so that the learnt representation encapsulates joint information between face and voice.
The output of the identity extractor is ingested by two additional fully-connected layers that are jointly trainable with the speech separation network.

In the speech encoder block, input speech signal is first transformed into a magnitude spectrum using short-time Fourier transform~(STFT), then the encoder network transforms the magnitude spectrum into a latent domain representation so that it is easy to combine with a visual representation extracted from the image encoder.

\vspace{2pt}
\noindent{\bf Audio-visual fusion.}
In the audio-visual fusion stage, we generate audio-visual joint features to represent speaker information as well as spectral information to be separated.
Joint audio-visual features are obtained by concatenating speech embeddings and visual embeddings along the channel axis.

We implement two strategies to assign visual identity on speech embeddings.
The first one uses equal speaker information on each frame of speech embeddings as done in~\cite{Wang2019voicefilter}.
The visual identity is concatenated on every speech embedding frames, after verifying its effectiveness in regulating the separation task by consistent guidance.
The other strategy is based on self-attention mechanism~\cite{lin2017selfatt} to provide differently weighted constraints for the separation.
Since each speech frame contains different states such as silence, target speech only, interfering speech only and overlapped speech, the degree of separation needs to be varied depending on the characteristic of each frame.
We extract weights referred to speech embeddings, and they are multiplied with the identity vector.
The weighted visual embedding vector is concatenated with its corresponding speech embedding frame.

\vspace{2pt}
\noindent{\bf Speech separation.}
The mask estimator predicts a time-frequency mask, which only leaves the desired voice from the observed signal.
In particular, it notices the target speaker using audio-visual features and generates a soft mask that considers the power ratio between target and interfering signals.
The magnitude spectrum is filtered with the estimated mask to remove undesired components including other speaker's voice.
Once the spectrogram is separated, the speech signal is synthesized using phase reconstruction and inverse STFT.
We use the Griffin-Lim algorithm~\cite{griffinlim1984} to reconstruct phase components because it is better than using the phase of input noisy signal, especially when the input signal is severely distorted by interfering speech.

\vspace{2pt}
\noindent{\bf Learning strategy.}
Our training strategy is described into two different criteria for speech separation $L_{SS}$ and speaker isolation $L_{SRL}$ respectively.
\begin{equation}
    L_{TOT} = L_{SS}+L_{SRL}
    \label{eq:totloss}
\end{equation}
The training criterion for the separation is based on minimising mean-squared-error~(MSE) between masked spectrogram and magnitude spectrum of reference speech in the logarithm scale~(Equation~\ref{eq:avseloss}).
\begin{equation}
    L_{SS} = \norm{log\bigg(\frac{X}{f(\vec{A}\doubleplus \vec{V})\odot S}\bigg)}^2
    \label{eq:avseloss}
\end{equation}
where $X$ is target magnitude spectrum, $S$ is mixed input speech spectrum and $f(\cdot),\vec{A},\vec{V}$ are the mask estimator, audio embeddings and weighted visual embeddings respectively, where $\doubleplus$ is concatenation mark. 
All network parameters are jointly trained except the face identity extractor in the image encoder.

In addition, we exploit another learning criterion, speaker representation loss~\cite{mun2020sound}, where it maximises the cosine similarity of latent distributions between the separated speech and the reference speech using the speech identity extractor pre-trained on cross-modal biometrics task (Equation~\ref{eq:spkloss}).
\begin{equation}
    L_{SRL} = 1 - \text{cos}\big(g(X), g(f(\vec{A}\doubleplus \vec{V})\odot S)\big)
    \label{eq:spkloss}
\end{equation}
where $g(\cdot)$ is the speech identity extractor.
It helps the model to isolate the target speaker from multiple speakers since the embeddings from the speech identity extractor represent speaker identity, hence only allowing the target speaker's identity to remain in separated signal.

\section{Experiments}
\label{sec:experiments}

\begin{figure*}[t]
    \centering
    \begin{minipage}[t]{0.48\linewidth}
            \centering
            \hspace{5pt}\begin{flushleft}\footnotesize{\fbox{ Male-Male}}\end{flushleft} \vspace{-5pt}
            \includegraphics[width=0.32\columnwidth]{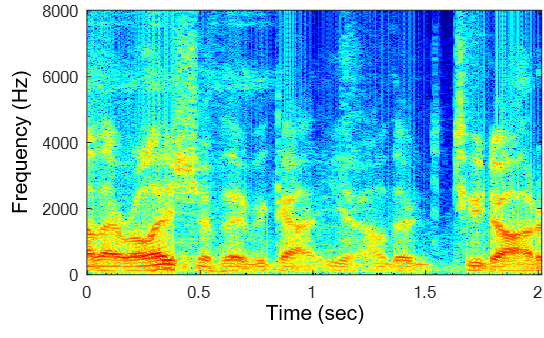}
            \includegraphics[width=0.32\columnwidth]{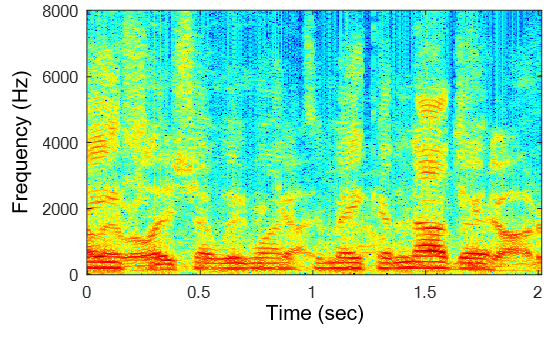}
            \includegraphics[width=0.32\columnwidth]{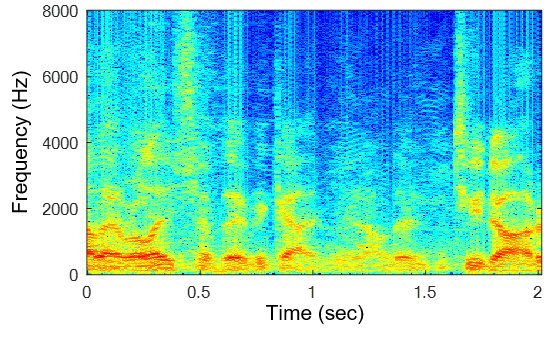}
    \end{minipage}
    \hspace{10pt}
    \begin{minipage}[t]{0.48\linewidth}
            \centering
            \hspace{5pt}\begin{flushleft}\footnotesize{\fbox{ Male-Female}}\end{flushleft} \vspace{-5pt}
            \includegraphics[width=0.32\columnwidth]{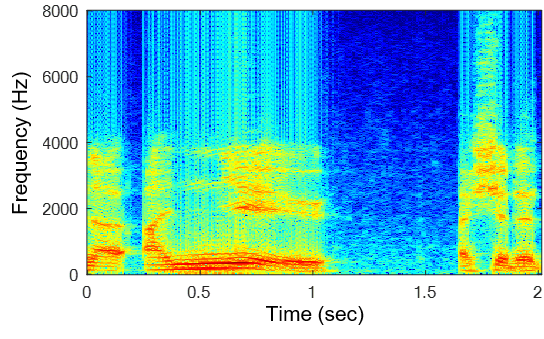}
            \includegraphics[width=0.32\columnwidth]{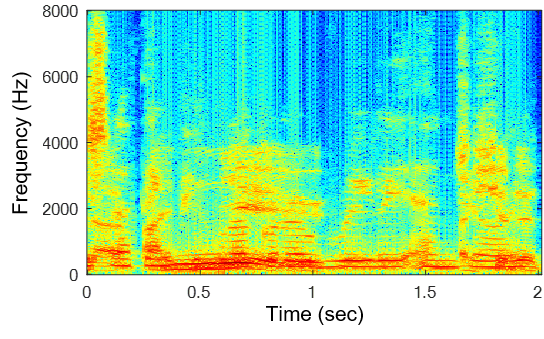}
            \includegraphics[width=0.32\columnwidth]{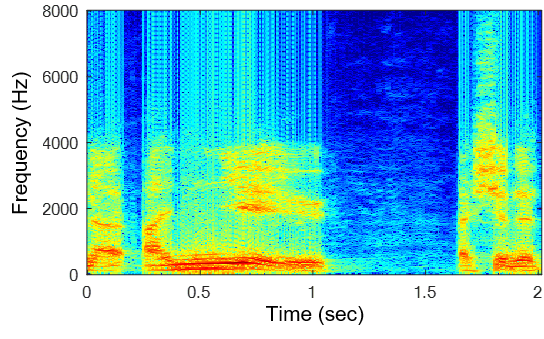}
    \end{minipage}

    \medskip \vspace{-15pt}
    
    \begin{minipage}[t]{0.48\linewidth}
            \centering
            \hspace{5pt}\begin{flushleft}\footnotesize{\fbox{ Female-Male}}\end{flushleft} \vspace{-5pt}
            \includegraphics[width=0.32\columnwidth]{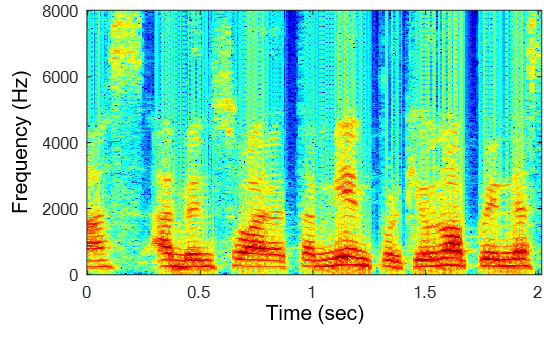}
            \includegraphics[width=0.32\columnwidth]{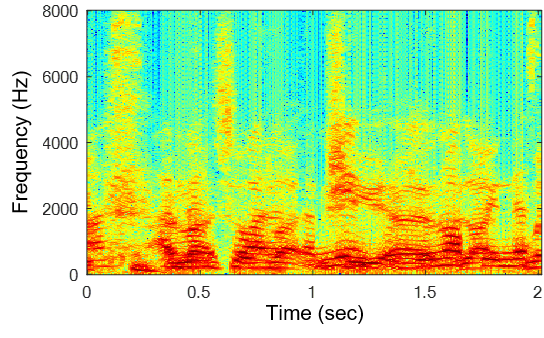}
            \includegraphics[width=0.32\columnwidth]{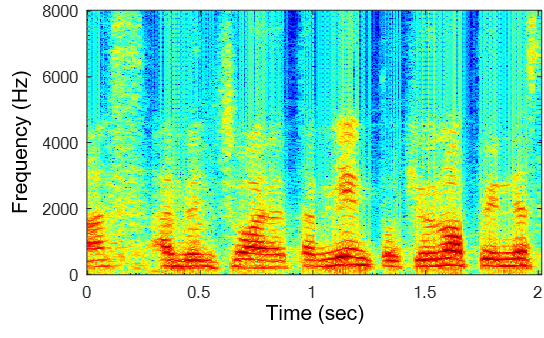}
    \end{minipage}
    \hspace{10pt}
    \begin{minipage}[t]{0.48\linewidth}
            \centering
            \hspace{5pt}\begin{flushleft}\footnotesize{\fbox{ Female-Female}}\end{flushleft} \vspace{-5pt}
            \includegraphics[width=0.32\columnwidth]{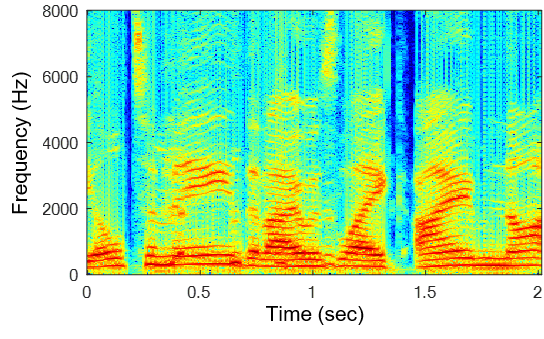}
            \includegraphics[width=0.32\columnwidth]{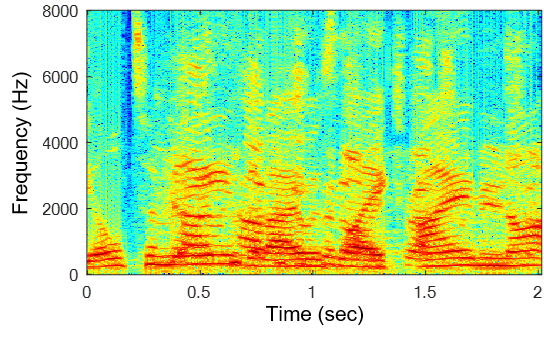}
            \includegraphics[width=0.32\columnwidth]{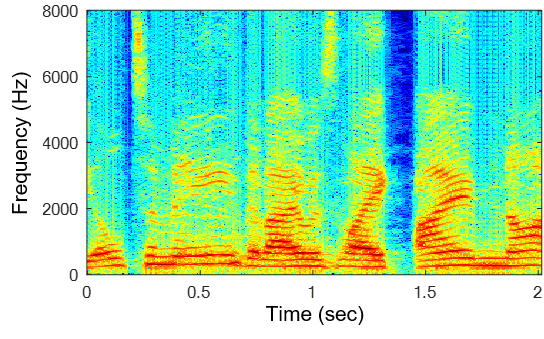}
    \end{minipage}
    \caption{Spectrogram for experimental results. There are 4 samples which are combinations of genders, where genders on the left are the targets and the right are set as interference. Each set demonstrates clean speech, mixed speech and separated speech from left to right respectively.}
    \vspace{-6pt}
\end{figure*}

\subsection{Experimental setup}
\label{subsec:settings}
Both the cross-modal biometrics model and the AVSS model are trained on the VoxCeleb2 dataset~\cite{Chung18b} which contains celebrity voices from YouTube.
The dataset contains 5,994 speakers with a total of 1,092,009 clips in the training set and 118 speakers with 36,237 clips in the test set.
Video clips in VoxCeleb2 include various environmental factors such as ambient noise, reverberation, and channel effects.

We prepare the training data for speech separation by mixing two clips from different speakers, while the protocol for generating the training data for the biometric task is identical to that in~\cite{Nagrani18}.
There are 2,000,000 samples for training and 100,000 samples for validation, both of which are from the training set speakers.
For evaluation, we reserve two types of data, which are {\em seen-heard} and {\em unseen-unheard} speakers, and each type has another 4 gender pair subsets, {\em i.e.,} male-male (\textit{M-M}), male-female (\textit{M-F}), female-male (\textit{F-M}), and female-female (\textit{F-F}).
Each evaluation set has 1,000 samples.
There is no overlapped evaluation data for the separation task with training set for the biometric task.

\vspace{2pt}
\noindent{\bf Audio representation.}
Audio stream of the cross-modal biometric network ingests 40-dimensional mel-spectrogram in logarithm scale, extracted at every 10ms with 25ms window length. 
Audio input to the speech separation module has same settings for slicing and shifting of frames, and input signal is transformed into 257 dimension using short-time Fourier transform.
Therefore, input dimension is $40\times200$ for the biometric network and $257\times200$ for the separation network, where the length of input audio segment is 2 seconds.

\vspace{2pt}
\noindent {\bf Image representation.}
The input to the visual network is a single image containing the person's face of which size is $224\times224$. 
For every video in VoxCeleb2, we sample 10 image frames, from which one image is randomly sampled during training.
The images are taken from the same video clip as the audio since the cross-modal training does not make use of the identity labels, and this also helps to prevent mismatch in person's physiological characteristics such as age.
For the evaluation, the images are randomly taken from other clips with same speaker, which is in line with the practical scenario where a profile image would be used.

\begin{table}[t]
    \centering
    \caption{Architecture of audio-visual speech separation network. The configurations of FCBlock, ConvBlock, and ResBlock are described in Figure~\ref{fig:block}.}\vspace{-5pt}
    \begin{tabular}{ c | l | r }
        \toprule
        {\bf Model}   & {\bf Layer} & {\bf \# Channels}\\
        \midrule\midrule
        \multirow{2}{*}{\makecell{FC layers \\ (Image encoder)}} & FCBlock & 1536 \\
                                        & FCBlock & 512 \\
        \midrule
        \multirow{1}{*}{Temporal attention} & FC + Sigmoid & 1 \\
        \midrule
        \multirow{3}{*}{\makecell{Encoder \\ (Speech encoder)}} & ConvBlock & 1536\\
                                        & ResBlock ($\times 4$)  & 1536\\
                                        & FCBlock & 512     \\
        \midrule
        \multirow{2}{*}{Mask estimator}     & ConvBlock ($\times 15$)  & 1536\\
                                            & FC + Sigmoid     & 257         \\
        \bottomrule
    \end{tabular}
    \label{table:arc_avse}
    \vspace{-6pt}
\end{table}

\begin{figure}[t]
    \centering
    \fbox{
    \begin{minipage}[b]{0.31\columnwidth}
        \centering
        \centerline{\includegraphics[width=0.9\columnwidth]{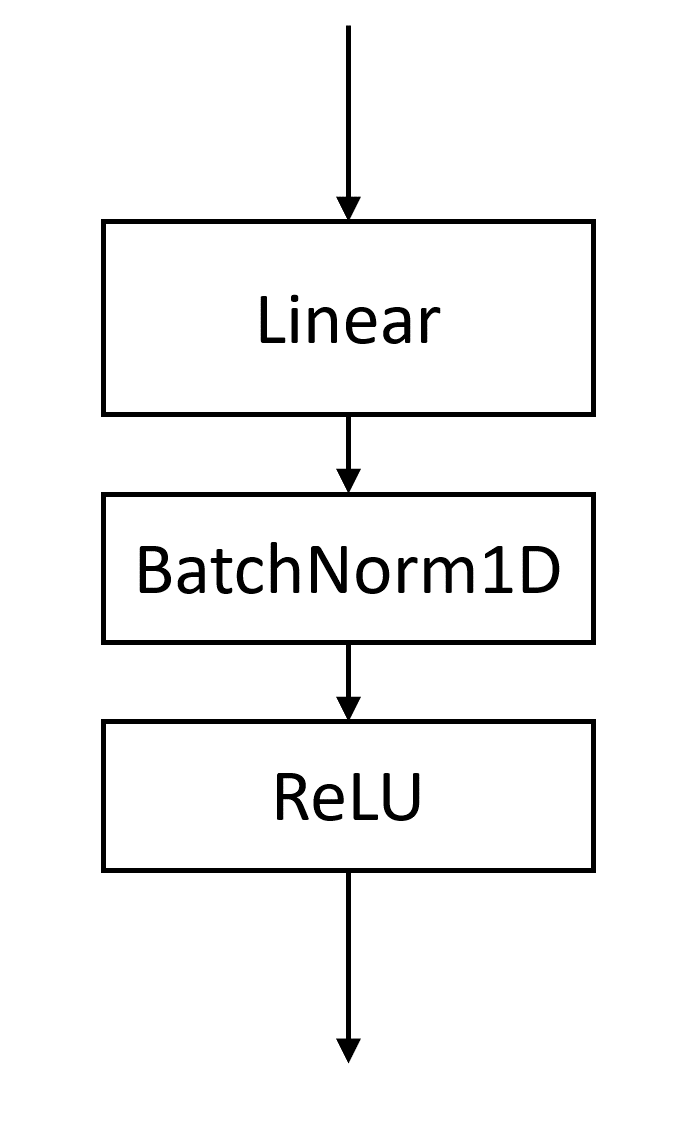}}
        \centerline{\footnotesize{(a) FCBlock}}\medskip
    \end{minipage}
    \begin{minipage}[b]{0.31\columnwidth}
        \centering
        \centerline{\includegraphics[width=0.9\columnwidth]{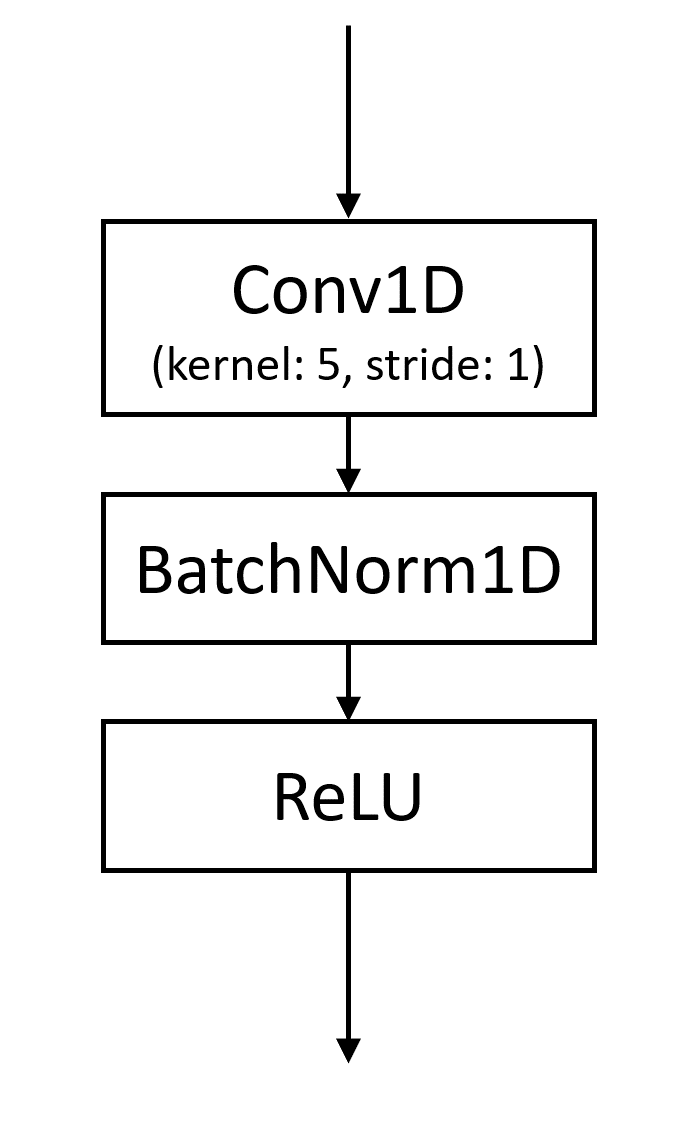}}
        \centerline{\footnotesize{(b) ConvBlock}}\medskip
    \end{minipage}
    \begin{minipage}[b]{0.31\columnwidth}
        \centering
        \centerline{\includegraphics[width=0.9\columnwidth]{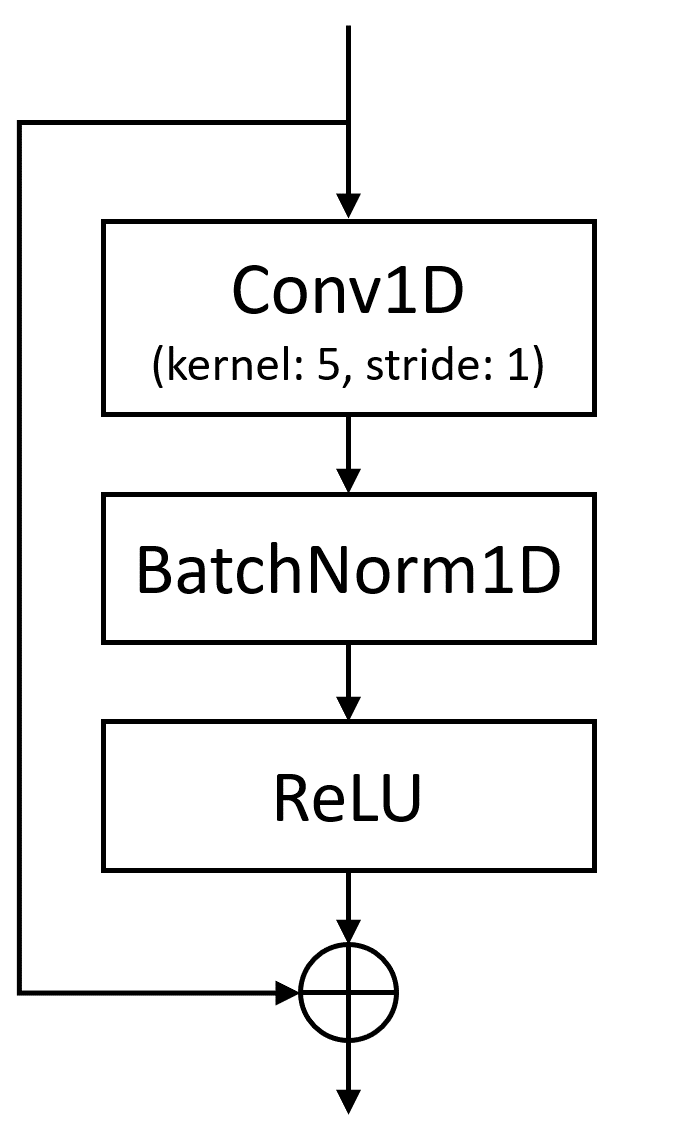}}
        \centerline{\footnotesize{(c) ResBlock}}\medskip
    \end{minipage}
    }
    \caption{Configuration of neural network blocks}
    \vspace{-5pt}
    \label{fig:block}
\end{figure}

\vspace{2pt}
\noindent {\bf Network parameters.}
Each sub-network of the proposed AVSS model mostly consists of convolutional layers.
The detailed parameter settings for the architecture are given in Table~\ref{table:arc_avse}, and the structures of {\em FCBlock}, {\em ConvBlock} and {\em ResBlock} are described in Figure~\ref{fig:block}.


\subsection{Evaluation protocol}
\label{subsec:evalprotocol}

The method is evaluated on two different tasks -- speech separation performance and speaker isolation. The first measures the separation performance, and signal-to-distortion~(SDR) is used to measure the quality of separated signals.
The secondary evaluation metric is speaker isolation accuracy which shows whether the output speech is from the target speaker given in the face image.
The accuracy indicates the effectiveness of visual information to indicate correct target speaker. 
Audio-only separation method (baseline) shows an accuracy of 49.94\% ({\em i.e.} random selection).

\begin{table}[t]
    \centering
    \caption{Audio-visual speech separation and speaker isolation results on VoxCeleb2}\vspace{-5pt}
    \begin{tabular}{l | l | r | r}
        \toprule
        {\bf Method}     & {\bf Criterion} & {\bf SDRi} & {\bf Accuracy} \\
        \midrule\midrule
        Audio-only  & MSE w. PIT           & 0.690 dB& 49.94 \%\\
        \midrule
        Audio-visual           & MSE       & 2.348 dB& 80.56 \%\\
        \hspace{5pt}+attention & MSE       & 2.413 dB& 80.89 \%\\
        \hspace{5pt}+attention & MSE w. SRL & 2.528 dB& 80.20 \%\\
        \bottomrule
    \end{tabular}
    \label{table:result_enh}
    \vspace{-6pt}
\end{table}    

\begin{table}[t]
    \centering
    \caption{Speech separation results for gender combinations}\vspace{-5pt}
    \begin{tabular}{L{2.5cm} | R{1.8cm} |  R{1.8cm}}
        \toprule
        {\bf Mixture} & \multicolumn{1}{r|}{\bf SDRi}  & {\bf Accuracy}\\
        \midrule\midrule
        \multicolumn{3}{c}{\bf Seen-Heard Speaker}\\ \midrule
        Male-Male       & 2.084 dB& 74.8 \%\\
        Male-Female     & 3.039 dB& 94.4 \%\\
        Female-Male     & 3.821 dB& 93.8 \%\\
        Female-Female   & 1.930 dB& 71.2 \%\\
        \midrule
        \multicolumn{3}{c}{\bf Unseen-Unheard Speaker}\\
        \midrule
        Male-Male       & 1.256 dB& 61.2 \%\\ %
        Male-Female     & 3.066 dB& 92.1 \%\\ %
        Female-Male     & 3.830 dB& 94.5 \%\\ %
        Female-Female   & 1.198 dB& 59.6 \%\\ %
        \bottomrule
    \end{tabular}
    \label{table:result_gender}
    \vspace{-6pt}
\end{table}    


\subsection{Experiment results}
Table~\ref{table:result_enh} reports the results of our proposed methods using the metrics described in Section~\ref{subsec:evalprotocol}.
Our baseline is a separation network trained using the permutation invariant training~(PIT) loss~\cite{yu2017pit}, a popular method for speech separation.
We compare three variants of the proposed method -- the basic AVSS model, the model with attention-based fusion, and the model trained with both $L_{SS}$ and $L_{SRL}$ losses.
The basic AVSS model does not include temporal attention and visual vector is padded uniformly along the time direction with the same values.

Note that the SDRi values are lower than those typically reported for source separation on other datasets, since clips in VoxCeleb2 contain environmental noise which degrades the SDR numbers when the network correctly removes the noise from the target.

We report accuracy for different gender pairs, which significantly affects the performance as shown in Table~\ref{table:result_gender}.
Mixtures with same gender, {\em i.e.,} \textit{M-M} and \textit{F-F} sets, are more difficult to distinguish by given the images, and the SDR improvement is relatively small.
The results for different gender sets, {\em i.e.,} \textit{F-M} and \textit{M-F} sets, show significantly higher SDR gain and identification accuracy since it is much easier to select the target speech using the given image.
Unseen-unheard test sets consist of samples that are not seen during the training of both cross-modal biometrics and speech separation tasks.
Although the discrimination performance of unseen people is slightly lower than that of seen pairs, the performance is still far above the baseline.

\section{Conclusion}
\label{sec:conclusion}

We proposed a novel audio-visual speech enhancement method that can isolate a specific speaker from multi-talker simultaneous speech using a conditional embedding represented by face image.
By using self-supervision, speaker representation can be retrieved from a face image in latent space, which is then used to condition the speech enhancement network.
This approach overcomes the permutation problem that is unavoidable in audio-only source separation, and consistently reconstructs speech from the target identity.
The experimental results confirm its effectiveness on the speech enhancement task.


\clearpage

\vfill
\newpage
\label{sec:refs}
\bibliographystyle{IEEEbib}
\bibliography{shortstrings,refs}

\end{document}